\documentclass[aps,preprint]{revtex4}%
\usepackage{amsfonts}
\usepackage{amsmath}
\usepackage{amssymb}
\usepackage{graphicx}%
\providecommand{\U}[1]{\protect\rule{.1in}{.1in}}
\newtheorem{theorem}{Theorem}
\newtheorem{acknowledgement}[theorem]{Acknowledgement}

\begin{document}
\preprint{ }
\title{Massive Gravity: Exorcising the Ghost\\ }
\author{Lasma Alberte}
\affiliation{Theoretical Physics, Ludwig Maxmillians University,Theresienstr. 37, 80333
Munich, Germany}
\author{Ali H. Chamseddine}
\affiliation{American University of Beirut, Physics Department, Beirut, Lebanon, and
I.H.E.S. F-91440 Bures-sur-Yvette, France}
\author{Viatcheslav Mukhanov}
\affiliation{Theoretical Physics, Ludwig Maxmillians University,Theresienstr. 37, 80333
Munich, Germany and Department of Physics, New York University, NY 10003, USA}
\pacs{PACS number}

\begin{abstract}
We consider Higgs massive gravity \cite{mukh,ACM} and investigate whether a
nonlinear ghost in this theory can be avoided. We show that although the
theory considered in \cite{gr1,gr} is ghost free in the decoupling limit, the
ghost nevertheless reappears in the fourth order away from the decoupling
limit. We also demonstrate that there is no direct relation between the value
of the Vainshtein scale and the existence of nonlinear ghost. We discuss how
massive gravity should be modified to avoid the appearance of the ghost.

\end{abstract}
\maketitle

\section{Introduction}

In \cite{mukh, ACM} we have devised a Higgs mechanism for massive gravity and
demonstrated how this theory goes smoothly to General Relativity below the
Vainshtein radius \cite{vain}, thus resolving the problem of van Dam, Veltman
and Zakharov discontinuity \cite{dam,zak}. This result, obtained in Higgs
massive gravity, is in agreement with the results derived in bigravity
theories in \cite{bab, bab2, bab3}. Moreover, we have found that the
corresponding Vainshtein scale depends on the nonlinear extension of the
Fierz-Pauli term \cite{pauli}. In particular, it was shown that the Vainshtein
scale can be changed within the range $M_{0}^{1/3}m_{g}^{-2/3}<R_{V}%
<M_{0}^{1/5}m_{g}^{-4/5},$ where $M_{0}$ and $m_{g}~$are, respectively, the
mass of the external source and the mass of the graviton in Planck units. The
class of actions which lead to different Vainshtein scales $R_{V}$ coincide
with the actions derived in \cite{gr1,gr}. These were obtained from the
requirement of absence of the nonlinear ghost \cite{boul} in the corresponding
order of perturbation theory, in the decoupling limit when both the graviton
mass and the gravitational constant simultaneously vanish, in such a way that
the appropriate Vainshtein scale is kept fixed. Moreover, there is a unique
action (up to total derivatives), corresponding to $R_{V}^{\infty}=M_{0}%
^{1/3}m_{g}^{-2/3},$ in the decoupling limit, for which the Boulware-Deser
ghost does not appear at all below Vainshtein energy scale, up to an arbitrary
order in perturbation theory \cite{gr1,gr}. Therefore, a natural interesting
question arises as to whether this result could be sustained if we consider
instead of the decoupling limit (which is not physical), the full nonlinear
theory of massive gravity. The answer to this question will also help us
understand whether there is any deep connection between the absence of
nonlinear ghost at a certain order in perturbation theory and the
corresponding value of the Vainshtein scale.

The main purpose of this note is to show that in the theories considered in
\cite{gr1,gr}, but away from the decoupling limit, the nonlinear ghost
inevitably arises in the fourth order of the perturbative expansion. The
Vainshtein scale value becomes therefore unrelated to the absence of ghost if
one does not consider the unrealistic decoupling limit of massive gravity.

The inevitable appearance of ghost in massive gravity theories agrees with an
independent argument of \cite{dv2} based on helicity decomposition.

We will also discuss how Higgs massive gravity must be modified if one wants
to avoid the appearance of the nonlinear ghost in any order of perturbative expansion.

\section{Higgs massive gravity}

We employ four scalar fields $\phi^{A},\,A=0,1,2,3$, to play the role of Higgs
fields. They will acquire a vacuum expectation value proportional to the
space-time coordinates $\phi^{A}=\delta_{\beta}^{A}x^{\beta}$ giving mass to
the graviton. Let us consider perturbations around Minkowski background,
\begin{equation}
g^{\mu\nu}=\eta^{\mu\nu}+h^{\mu\nu},\text{ \ \ \ \ \ }\phi^{A}=x^{A}+\chi
^{A}\label{1n}%
\end{equation}
and define
\begin{align}
\bar{h}_{B}^{A} &  \equiv\eta_{BC}g^{\mu\nu}\partial_{\mu}\phi^{A}%
\partial_{\nu}\phi^{C}-\delta_{B}^{A}\nonumber\\
&  =h_{B}^{A}+\partial^{A}\chi_{B}+\partial_{B}\chi^{A}+\partial_{C}\chi
^{A}\partial^{C}\chi_{B}\nonumber\\
&  \ +h_{C}^{A}\partial^{C}\chi_{B}+h_{B}^{C}\partial_{C}\chi^{A}+h_{D}%
^{C}\partial^{D}\chi_{B}\partial_{C}\chi^{A},\label{2n}%
\end{align}
where indices are moved with the Minkowski metric $\eta_{AB}=(1,-1,-1,-1)$, in
particular, $\chi_{B}=\eta_{BC}\chi^{C}$ and $h_{B}^{A}=\eta_{BC}\delta_{\mu
}^{A}\delta_{\nu}^{C}h^{\mu\nu}.$ After introducing the diffeomorphism
invariant variable $\bar{h}_{B}^{A}$ it becomes almost trivial to write the
terms that produce massive gravity. In the unitary gauge where $\chi^{A}=0,$
we have $\bar{h}_{B}^{A}=h_{B}^{A}=\eta_{BC}\delta_{\mu}^{A}\delta_{\nu}%
^{C}h^{\mu\nu}$, and hence the Fierz-Pauli term for the graviton mass around
broken symmetry background can immediately be obtained from the quadratic term
of the following action for the scalar fields%
\begin{equation}
S_{\phi}=\frac{m_{g}^{2}}{8}\int d^{4}x\,\sqrt{-g}\left[  \bar{h}^{2}-\bar
{h}_{B}^{A}\bar{h}_{A}^{B}+O\left(  \bar{h}^{3},...\right)  \right]
.\label{3n}%
\end{equation}
where by $O\left(  \bar{h}^{3},...\right)  $ we denote the terms which are of
third and higher orders in $\bar{h}_{B}^{A}.$ In distinction from the
Fierz-Pauli action which was introduced by explicit spoiling of diffeomorphism
invariance, our action is manifestly diffeomorphism invariant and only
coincides, to leading order, with the Fierz-Pauli action, in the unitary gauge
where all perturbations of the scalar fields are set to zero.

\section{Boulware-Deser nonlinear ghost}

One could, in principle, skip all higher order terms and consider the action
\begin{equation}
S=-\frac{1}{2}\int d^{4}x\,\sqrt{-g}R+\frac{m_{g}^{2}}{8}\int d^{4}%
x\,\sqrt{-g}\left[  \bar{h}^{2}-\bar{h}_{B}^{A}\bar{h}_{A}^{B}\right]
,\label{3nb}%
\end{equation}
where we set $8\pi G=1,$ as an exact action for massive gravity. The problem
then is either the presence of a ghost around the trivial background $\phi
^{A}=0$ or the appearance of nonlinear ghost in the broken symmetry phase. To
trace the latter one it is convenient to work in some gauge where the scalar
field perturbations are not equal to zero. A good choice is the Newtonian
gauge in which the metric $g_{\mu\nu}$ takes the form \cite{vm}%
\begin{equation}
ds^{2}=\left(  1+2\phi\right)  dt^{2}+2S_{i}dtdx^{i}-\left[  \left(
1-2\psi\right)  \delta_{ik}+\tilde{h}_{ik}\right]  dx^{i}dx^{k},\label{3na}%
\end{equation}
where $S_{i,i}=0$ and $\tilde{h}_{ij,i}=\tilde{h}_{ii}=0.$ Then the ghost can
easily be traced as a dynamical degree of freedom of the scalar field
$\chi^{0}.$ The field $\chi^{0}$ enters only the $\bar{h}_{0}^{0}$ and
$\bar{h}_{0}^{i}$ components, which can be written explicitly as%
\begin{align}
\bar{h}_{0}^{0} &  =h^{00}+2\dot{\chi}^{0}+2h^{00}\dot{\chi}^{0}+\left(
\dot{\chi}^{0}\right)  ^{2}+h^{00}\left(  \dot{\chi}^{0}\right)  ^{2}%
+2h^{0i}\dot{\chi}^{0}\chi_{,i}^{0}\label{4n}\\
&  +2h^{oi}\partial_{i}\chi^{0}-\delta^{ik}\chi_{,i}^{0}\chi_{,k}^{0}%
+h^{ik}\chi_{,i}^{0}\chi_{,k}^{0}\nonumber
\end{align}
and
\begin{align}
\bar{h}_{0}^{i} &  =h^{0i}+\dot{\chi}^{i}+h^{00}\dot{\chi}^{i}-\delta^{ik}%
\chi_{,k}^{0}+h^{ik}\chi_{,k}^{0}+\left(  h^{0i}+\dot{\chi}^{i}+h^{00}%
\dot{\chi}^{i}+h^{k0}\chi_{,k}^{i}\right)  \dot{\chi}^{0}\nonumber\\
&  +h^{k0}\chi_{,k}^{i}-\delta^{lk}\chi_{,l}^{i}\chi_{,k}^{0}+h^{lk}\chi
_{,l}^{i}\chi_{,k}^{0}+h^{0k}\chi_{,k}^{0}\dot{\chi}^{i}.\label{5n}%
\end{align}
Let us consider only the scalar mode of the massive graviton for which
$\chi^{i}=\pi_{,i}$. It was shown in \cite{ACM} that by using constraints one
can express the linear perturbations of the scalar fields in terms of the
metric potential $\psi$ as%
\begin{align}
\pi &  =\frac{2\Delta-3m_{g}^{2}}{m_{g}^{2}\Delta}\psi\label{6nb}\\
\chi^{0} &  =-\frac{2\Delta+3m_{g}^{2}}{m_{g}^{2}\Delta}\dot{\psi}.\label{6na}%
\end{align}
Then the action (\ref{3nb}) up to second order in perturbations simplifies to%
\begin{equation}
^{(S)}\delta_{2}S=-3\int d^{4}x\,\left[  \psi\left(  \partial_{t}^{2}%
-\Delta+m_{g}^{2}\right)  \psi\right]  .\label{7n}%
\end{equation}
The nonlinear ghost appears in the third order in metric and scalar field
perturbations. This is due to the fact that the accidental $U(1)$ symmetry,
which makes the scalar field $\chi^{0}$ to be the Lagrange multiplier around
Minkowski background, is not preserved on a background slightly deviating from
Minkowski space \cite{mukh}. To prove this it is enough to consider only the
third order terms in the action (\ref{3nb}) which involve the powers of
$\dot{\chi}^{0}.$ By substituting (\ref{4n}) and (\ref{5n}) into (\ref{3nb})
we obtain%
\begin{equation}
\delta_{3}S=\frac{m_{g}^{2}}{2}\int d^{4}x\left\{  \left[  \left(
h^{00}+\delta\sqrt{-g}\right)  \bar{h}_{i}^{i}+\left(  h^{0i}+\dot{\chi}%
^{i}-\chi_{,i}^{0}\right)  \left(  h^{0i}+\dot{\chi}^{i}\right)  \right]
\dot{\chi}^{0}+\frac{1}{2}\bar{h}_{i}^{i}\left(  \dot{\chi}^{0}\right)
^{2}+...\right\}  ,\label{8n}%
\end{equation}
where by dots we have denoted all other terms not containing time derivatives
of $\chi^{0}.$ The term, linear in $\dot{\chi}^{0},$ does not induce dynamics
for the mode $\chi^{0}$ and simply modifies the constraint equations to second
order in perturbations. However, the term proportional to $\left(  \dot{\chi
}^{0}\right)  ^{2}$ induces the propagation of $\chi^{0}$ on the background
deviating from Minkowski space for which $\bar{h}_{i}^{i}\neq0.$ Thus at
nonlinear level there appears an extra scalar degree of freedom which is a
ghost. To see this let us express the relevant term in (\ref{8n}) entirely in
terms of the gravitational potential $\psi.$ Taking into account that, to
linear order, $\bar{h}_{i}^{i}=6\psi+2\Delta\pi$ and using constraint
equations (\ref{6nb}) and (\ref{6na}) we find%
\begin{equation}
\delta_{3}S=\frac{m_{g}^{2}}{4}\int d^{4}x\left[  \bar{h}_{i}^{i}\left(
\dot{\chi}^{0}\right)  ^{2}+...\right]  =\int d^{4}x\left[  \Delta\psi\left(
\frac{2\Delta+3m_{g}^{2}}{m_{g}^{2}\Delta}\ddot{\psi}\right)  ^{2}+...\right]
.\label{9n}%
\end{equation}
By considering inhomogeneities with $\Delta\psi\gg m_{g}^{2}\psi$ and
combining this contribution to the action (\ref{7n}) we obtain%
\begin{equation}
\delta S=-3\int d^{4}x\,\left[  \psi\left(  \partial_{t}^{2}-\Delta+m_{g}%
^{2}\right)  \psi-\frac{4}{3m_{g}^{4}}\Delta\psi\left(  \ddot{\psi}\right)
^{2}+...\right]  .\label{9nn}%
\end{equation}
Let us assume that there is a background field $\psi_{b}$ and consider small
perturbations around this background, that is, $\psi=\psi_{b}+\delta\psi.$
Expanding (\ref{9nn}) to second order in $\delta\psi$ we find that the
behavior of linear perturbations is determined by the action%
\begin{equation}
\delta S=-3\int d^{4}x\,\left\{  \delta\psi\left(  \partial_{t}^{2}%
-\Delta+m_{g}^{2}\right)  \delta\psi+\frac{1}{m_{Gh}^{2}}\left[  \left(
\partial_{t}^{2}\delta\psi\right)  ^{2}+2\frac{\ddot{\psi}_{b}}{\Delta\psi
_{b}}\left(  \Delta\delta\psi\right)  \left(  \partial_{t}^{2}\delta
\psi\right)  \right]  +...\right\}  ,\label{9nnn}%
\end{equation}
where%
\begin{equation}
m_{Gh}^{2}=-\frac{3m_{g}^{4}}{4\Delta\psi_{b}},\label{9nba}%
\end{equation}
Let us take for the background field the scalar mode of gravitational wave
with the wave-number $k\sim m_{g},$ for which $\ddot{\psi}_{b}\sim\Delta
\psi_{b}\sim m_{g}^{2}\psi_{b}$ and $m_{Gh}^{2}\sim m_{g}^{2}/\psi_{b}$. By
considering perturbations $\delta\psi$ with wave-numbers $m_{Gh}^{2}\gg
k^{2}\gg m_{g}^{2},$ and skipping subdominant terms, we can rewrite the action
above as%
\begin{equation}
\delta S\approx-\frac{3}{m_{Gh}^{2}}\int d^{4}x\delta\psi\left(  \partial
_{t}^{2}+...\right)  \left(  \partial_{t}^{2}+m_{Gh}^{2}+...\right)
\delta\psi.\label{10na}%
\end{equation}
The perturbation propagator is then given by
\begin{equation}
\frac{1}{\partial^{2}\left(  \partial^{2}+m_{Gh}^{2}\right)  }\simeq\frac
{1}{m_{Gh}^{2}}\left(  \frac{1}{\partial^{2}}-\frac{1}{\partial^{2}+m_{Gh}%
^{2}}\right)  ,\label{12n}%
\end{equation}
and it obviously describes the scalar mode of the graviton together with
non-perturbative Boulware-Deser ghost of mass $m_{Gh}\sim m_{g}/\sqrt{\psi
_{b}}.$ It is clear that when $\psi_{b}$ vanishes the mass $m_{Gh}$ becomes
infinite and ghost disappears. We have argued in \cite{ACM} that at energies
above Vainshtein scale $\Lambda_{5}=m_{g}^{4/5}$ the linearized consideration
above breaks down and the scalar fields enter the strong coupling regime.
Therefore, if $m_{Gh}$ would be larger than $\Lambda_{5}$ then this ghost
would not be essential. However, in strong enough background $m_{g}%
<m_{Gh}<\Lambda_{5}$ and therefore the nonlinear ghost appears below the
Vainshtein scale where it is visible.

Thus, the action (\ref{3nb}) considered as describing massive gravity has two
problems with ghosts: first, there is a linear ghost around the trivial
background $\phi^{A}=0$, and second, there is nonlinear ghost around broken
symmetry background.

The first ghost is dangerous, because it leads to a strong instability.
However, as we have shown in \cite{mukh}, it can be easily avoided by adding
to the action (\ref{3nb}) third and higher order terms in $\bar{h}.$ This
modification is ambiguous and there is a whole class of theories which
reproduce the Fierz-Pauli theory in the lowest order, avoiding linear ghosts
around trivial background.

The nonlinear ghost exists only at scales below the Vainshtein energy scale
which, for the realistic graviton mass, is extremely low, about $10^{-20}eV$.
Therefore, taking into account that the Vainshtein scale serves as the cutoff
scale in Lorentz violating background, where the nonlinear ghost propagates,
we conclude that this ghost is completely harmless in agreement with
\cite{CJM}. Nevertheless, some interesting questions remain. One could inquire
whether there is any nonlinear extension of the action (\ref{3nb}) which is
free of the Boulware-Deser ghost and how the absence of the ghost in the
corresponding order of a perturbative expansion is related with the concrete
value of the Vainshtein scale?

\section{Ghost in nonlinear extensions of massive gravity}

Contrary to \cite{ddgv,crim,dhk}, it was claimed recently in \cite{gr1,gr},
that there is unique ghost-free nonlinear extension of massive gravity and
that this extension is related with $\Lambda_{3}=m_{g}^{2/3}$ Vainshtein
scale. This claim was proved in \cite{gr1,gr} in the decoupling limit
neglecting the vector modes of the graviton. The decoupling limit, while
simplifying the calculations, is not physically justified. Therefore, we will
determine whether the nonlinear ghost really disappears away from the
decoupling limit. The Lagrangian in \cite{gr1,gr} is expressed in terms of the
invariants built out of%
\begin{equation}
H_{\mu\nu}=g_{\mu\nu}-\eta_{AB}\partial_{\mu}\phi^{A}\partial_{\nu}\phi
^{B}.\label{13n}%
\end{equation}
It is easy to see (as was also noted in \cite{bere}) that the invariants built
out of $H_{\mu\nu},$ up to sign, coincide with the invariants made of $\bar
{h}_{B}^{A},$ in particular,%
\begin{equation}
g^{\mu\nu}H_{\mu\nu}=-\bar{h},\text{ \ \ }H_{\mu\nu}H^{\mu\nu}=\bar{h}_{B}%
^{A}\bar{h}_{A}^{B},\text{ \ \ }...\label{14n}%
\end{equation}
Let us consider the action \cite{gr1,gr}:%
\begin{align}
S_{\phi} &  =\frac{m_{g}^{2}}{8}\int d^{4}x\,\sqrt{-g}\left[  \bar{h}^{2}%
-\bar{h}_{AB}^{2}+\frac{1}{2}\left(  \bar{h}_{AB}^{3}-\bar{h}\bar{h}_{AB}%
^{2}\right)  -\frac{5}{16}\bar{h}_{AB}^{4}+\frac{1}{4}\bar{h}\bar{h}_{AB}%
^{3}+\frac{1}{16}\left(  \bar{h}_{AB}^{2}\right)  ^{2}\right.  \nonumber\\
&  \left.  +c_{3}\left(  2\bar{h}_{AB}^{3}-3\bar{h}\bar{h}_{AB}^{2}+\bar
{h}^{3}+\frac{3}{4}\left(  2\bar{h}_{AB}^{3}\bar{h}-2\bar{h}_{AB}^{4}+\left(
\bar{h}_{AB}^{2}\right)  ^{2}-\bar{h}_{AB}^{2}\bar{h}^{2}\right)  \right)
\right.  \nonumber\\
&  \left.  +d_{5}\left(  6\bar{h}_{AB}^{4}-8\bar{h}_{AB}^{3}\bar{h}-3\left(
\bar{h}_{AB}^{2}\right)  ^{2}+6\bar{h}_{AB}^{2}\bar{h}^{2}-\bar{h}^{4}\right)
\right]  ,\label{14nn}%
\end{align}
where $c_{3}$ and $d_{5}$ are arbitrary coefficients and we have introduced
the shortcut notations
\[
\bar{h}_{AB}^{2}=\bar{h}_{B}^{A}\bar{h}_{A}^{B},\text{ \ }\bar{h}_{AB}%
^{3}=\bar{h}_{B}^{A}\bar{h}_{C}^{B}\bar{h}_{A}^{C},\text{ \ }\bar{h}_{AB}%
^{4}=\bar{h}_{B}^{A}\bar{h}_{C}^{B}\bar{h}_{D}^{C}\bar{h}_{A}^{D}.
\]
It was proved \cite{gr1,gr} that this theory is ghost free to fourth order in
perturbations in the decoupling limit. The action above corresponds to the
Vainshtein scale $\Lambda=m_{g}^{8/11}$ \cite{ACM}. Let us investigate whether
the ghost really disappears in non-decoupling limit. For this purpose we have
to trace all fourth order terms in perturbations which contain time
derivatives of $\chi^{0}.$ As we have noticed above, the time derivatives of
$\chi^{0}$ come only from $\bar{h}_{0}^{0}$ and $\bar{h}_{0}^{i}$ components.
Therefore the only terms in (\ref{14nn}), which survive and could be relevant
for a possible ghost are the following%
\begin{align}
S_{\phi} &  =\frac{m_{g}^{2}}{8}\int d^{4}x\,\sqrt{-g}\left[  \left(  2\bar
{h}_{0}^{0}-\frac{1}{2}\left(  \bar{h}_{0}^{0}\right)  ^{2}+\frac{1}{4}\left(
\bar{h}_{0}^{0}\right)  ^{3}\right)  \bar{h}_{i}^{i}-\frac{1}{4}\left(
2\bar{h}_{0}^{0}-\frac{1}{2}\left(  \bar{h}_{0}^{0}\right)  ^{2}\right)
\bar{h}_{ik}^{2}\right.  \nonumber\\
&  \left.  +2\bar{h}_{0}^{i}\bar{h}_{0}^{i}-\frac{1}{2}\bar{h}_{0}^{0}\bar
{h}_{0}^{i}\bar{h}_{0}^{i}+\frac{1}{4}\left(  \bar{h}_{0}^{0}\right)  ^{2}%
\bar{h}_{0}^{i}\bar{h}_{0}^{i}+\frac{3}{2}c_{3}\left(  2\bar{h}_{0}^{0}%
-\frac{1}{2}\left(  \bar{h}_{0}^{0}\right)  ^{2}\right)  \left(  \left(
\bar{h}_{i}^{i}\right)  ^{2}-\bar{h}_{ik}^{2}\right)  +...\right]
.\label{15nnn}%
\end{align}
We have skipped here the terms which are linear in $\dot{\chi}^{0}$ because
they only modify the constraints without inducing the dynamics for $\chi^{0}.$
We would like to stress that the particular choice of action (\ref{14nn}) has
lead to nontrivial cancelations of many terms which could have caused the
appearance of a ghost. In particular, all contributions which induce the terms
proportional to $\left(  \dot{\chi}^{0}\right)  ^{2},$ $\left(  \dot{\chi}%
^{0}\right)  ^{3},$ $\left(  \dot{\chi}^{0}\right)  ^{4}$ are cancelled in the
$d_{5}$ term in (\ref{14nn}). Further nontrivial cancelations happen when we
substitute (\ref{4n}) and (\ref{5n}) in (\ref{15nnn}), and the final result is%
\begin{equation}
\delta_{3}S_{\phi}+\delta_{4}S_{\phi}=\frac{m_{g}^{2}}{8}\int d^{4}x\left[
F\left(  \delta g,\chi\right)  \dot{\chi}^{0}+\frac{1}{2}\left(  \dot{\chi
}^{i}+g^{0i}+\chi_{,i}^{0}\right)  ^{2}\left(  \dot{\chi}^{0}\right)
^{2}+...\right]  ,\label{16n}%
\end{equation}
where we denote by dots the terms which do not depend on $\dot{\chi}^{0}.$
Note that the the third and fourth powers of $\dot{\chi}^{0}$ are canceled.
The function $F\left(  \delta g,\chi\right)  $ is some rather long and
complicated expression which depend on terms of second and third order in
perturbations but does not depend on $\dot{\chi}^{0}.$ Because this term does
not induce the dynamics of $\chi^{0},$ but simply modifies the constraints, we
do not need the explicit form of $F.$ Note that the third order terms with
second and third powers of $\dot{\chi}^{0}$ are canceled and hence the ghost
does not appear in the third order even if we do not consider the decoupling
limit \cite{bere}. However, in the fourth order in perturbations the nonlinear
ghost survives. It is easy to see that this ghost disappears in the decoupling
limit in agreement with \cite{gr,gr1,bere}. In fact, after skipping the vector
modes, we have $\chi^{i}=\pi_{,i}$ , $S_{i}=0$ and considering the decoupling
limit ($m_{g}^{2}\rightarrow0$) we obtain from (\ref{6nb}) and (\ref{6na})
that $\chi^{0}\rightarrow-\dot{\pi}$ and hence the second term in (\ref{16n})
vanishes. However, without taking this limit, action (\ref{16n}) becomes%
\begin{align}
\delta_{3}S_{\phi}+\delta_{4}S_{\phi} &  =\frac{m_{g}^{2}}{16}\int
d^{4}x\left[  \left(  \dot{\tilde{\chi}}^{i}+S_{i}+\left(  \dot{\pi}+\chi
^{0}\right)  _{,i}\right)  ^{2}\left(  \dot{\chi}^{0}\right)  ^{2}+...\right]
\nonumber\\
&  =\frac{m_{g}^{2}}{16}\int d^{4}x\left[  \left(  \dot{\tilde{\chi}}%
^{i}+S_{i}-\frac{6}{\Delta}\dot{\psi}_{,i}\right)  ^{2}\left(  \frac
{2\Delta+3m_{g}^{2}}{m_{g}^{2}\Delta}\ddot{\psi}\right)  ^{2}+...\right]
\label{18n}%
\end{align}
where we have taken into account that $\chi^{i}=\pi_{,i}+\tilde{\chi}^{i}$ and
$\tilde{\chi}^{i}$ is a vector mode of the graviton. Considering small
perturbations $\delta\psi$ with wave-numbers $k^{2}\gg m_{g}^{2}$ around some
background $\psi_{b}$ and $\tilde{\chi}_{b}^{i}$ we find as in the previous
considerations (see (\ref{9nn})-(\ref{9nba})) that this action describes,
along with the scalar mode of graviton, also a ghost of mass%
\begin{equation}
m_{Gh}^{2}=-12m_{g}^{2}\left(  \dot{\tilde{\chi}}_{b}^{i}+S_{i}-\frac
{6}{\Delta}\dot{\psi}_{b,i}\right)  ^{-2}\label{20n}%
\end{equation}
provided that $m_{Gh}^{2}$ satisfies the condition $\partial_{t}^{2}%
m_{Gh}^{-2}\ll1.$ In the background of the scalar gravitational wave $\psi
_{b}$ with $k^{2}\simeq m_{g}^{2}$ we have $m_{Gh}\sim m_{g}/\psi_{b}.$ If the
time dependent background fields are strong enough the mass of this ghost is
smaller than the Vainshtein scale and can be even as small as the graviton
mass. Thus, if one does not consider the decoupling limit of the theory the
action (\ref{14nn}) has a nonlinear ghost in the fourth order of perturbation
theory. This ghost cannot be removed by adding fifth and higher order terms
and it is inevitable in the theories considered in \cite{gr1,gr}.

\section{Can we avoid a nonlinear ghost?}

The theory described by action (\ref{14nn}) could be a unique candidate for a
ghost free massive gravity (to fourth order in perturbations) because it is
the only theory which does not have a ghost in the decoupling limit
\cite{gr1,gr}. Its higher order extension which removes ghost to an arbitrary
order is also uniquely determined by the requirement of the absence of ghost
in decoupling limit. Thus the theory satisfies the necessary condition to be a
ghost free theory. However, this condition is not sufficient to avoid ghost
when away from the non-realistic decoupling limit. Unfortunately, as we have
shown, the theory considered above inevitably has unremovable nonlinear ghost
beginning with the fourth order in perturbations. One can wonder whether there
is any way of avoiding this no-go theorem? It is clear that using $H_{\mu\nu}$
defined in (\ref{13n}) one is forced to use only the invariants present in
(\ref{14nn}) because otherwise the fundamental diffeomorphism invariance of
the theory will be spoiled. On the other hand in our approach we are not
obliged to preserve the fake \textquotedblleft Lorentz
invariance\textquotedblright\ in the space of scalar field configurations,
which was used to reproduce the Fierz-Pauli term. In fact, there is nothing
wrong from the point of view of symmetries to consider for instance the
Lagrangian%
\begin{equation}
S_{\phi}=\frac{m_{g}^{2}}{8}\int d^{4}x\,\sqrt{-g}\left(  g^{\mu\nu}%
\partial_{\mu}\phi^{0}\partial_{\nu}\phi^{0}-1\right)  ^{2}=\frac{m_{g}^{2}%
}{8}\int d^{4}x\,\sqrt{-g}\left(  \bar{h}_{0}^{0}\right)  ^{2}\label{15nna}%
\end{equation}
which is diffeomorphism and Lorentz invariant and simply describes the scalar
field $\phi^{0}$ with unusual kinetic term. Therefore, without spoiling any
fundamental invariance we could modify the action above by adding to it terms
of the form $\left(  \bar{h}_{i}^{0}\right)  ^{2}\bar{h}_{0}^{0},$ $\left(
\bar{h}_{i}^{0}\right)  ^{2},$ etcetera. It is easy to verify that the only
terms in (\ref{15nnn}) responsible for ghost are
\begin{equation}
\delta S_{Ghost}\equiv\frac{m_{g}^{2}}{8}\int d^{4}x\left[  2\bar{h}_{0}%
^{i}\bar{h}_{0}^{i}-\frac{1}{2}\bar{h}_{0}^{i}\bar{h}_{0}^{i}\bar{h}_{0}%
^{0}+\frac{1}{4}\bar{h}_{0}^{i}\bar{h}_{0}^{i}\left(  \bar{h}_{0}^{0}\right)
^{2}\right]  .\label{15o}%
\end{equation}
Therefore subtracting these terms from action (\ref{14nn}) removes the ghost
in the fourth order. In turn this also inevitably modifies the quadratic part
of the action and instead of Fierz-Pauli term we obtain%
\begin{align}
S_{\phi} &  =\frac{m_{g}^{2}}{8}\int d^{4}x\,\sqrt{-g}\left[  \bar{h}^{2}%
-\bar{h}_{B}^{A}\bar{h}_{A}^{B}-2\bar{h}_{0}^{i}\bar{h}_{0}^{i}+O\left(
\bar{h}^{3},...\right)  \right]  \nonumber\\
&  =\frac{m_{g}^{2}}{8}\int d^{4}x\,\sqrt{-g}\left[  \left(  \bar{h}_{i}%
^{i}\right)  ^{2}-\bar{h}_{k}^{i}\bar{h}_{i}^{k}+2\bar{h}_{0}^{0}\bar{h}%
_{i}^{i}+O\left(  \bar{h}^{3},...\right)  \right]  .\label{16a}%
\end{align}
As a result both scalar and vector modes of the graviton disappear and the
action above describes the massive transverse graviton with two degrees of
freedom. Note that this result does not contradict Wigner's theorem about the
number of degrees of freedom of massive particle with spin-two because in this
case the scalar fields background in the broken symmetry phase in not Lorentz
invariant. Nevertheless, we would like to stress that in Higgs gravity which
produces the massive graviton with two degrees of freedom there is no
violation of fundamental space-time Lorentz invariance (compare to
\cite{rubakov,dub}). Its effective violation is simply due to the existence of
a background scalar field in Minkowski space in a way similar to the violation
of this invariance by the cosmic microwave background radiation in our
universe. In the case when we have imposed the extra \textquotedblleft Lorentz
invariance\textquotedblright\ in the configuration space of the scalar fields
we were able to imitate the space-time Lorentz invariance for the graviton
mass term simply via redefinition of the scalar fields. However, in general
when this invariance is absent any scalar fields background violates
space-time Lorentz invariance explicitly.

The \textquotedblleft Lorentz violating\textquotedblright\ procedure of
removing the nonlinear ghost in Higgs gravity can be extended to any higher
orders in the theory considered in \cite{gr1,gr}. However, if we allow the
\textquotedblleft Lorentz violating\textquotedblright\ terms then there is no
need anymore for such extension. We can simply consider
\begin{equation}
S_{\phi}=\frac{m_{g}^{2}}{8}\int d^{4}x\,\sqrt{-g}\left[  \left(  \bar{h}%
_{i}^{i}\right)  ^{2}-\bar{h}_{k}^{i}\bar{h}_{i}^{k}\right]  ,\label{17n}%
\end{equation}
as an exact action of massive gravity on a Lorentz violating background. It is
obvious that this action depends only on three scalar fields and does not have
any linear and nonlinear ghosts around any background. The transverse
gravitational degrees of freedom $\tilde{h}_{ik}$ become massive and one could
wonder how it will modify the usual Newtonian interaction between massive
objects. To answer this question let us consider a static gravitational field
produced by a matter for which only $T^{00}$ component of the energy-momentum
tensor does not vanish. The metric in this case can be written as
\begin{equation}
ds^{2}=\left(  1+2\phi\right)  dt^{2}-\left(  1-2\psi\right)  \delta
_{ik}dx^{i}dx^{k},\label{18nn}%
\end{equation}
and the action for static perturbations derived in \cite{ACM} (see formulae
(28) and (36) there) in the case of (\ref{17n}) simplifies to%
\begin{align}
^{(S)}\delta S &  =\int d^{4}x\,\left\{  \psi_{,i}\psi_{,i}+\phi\left[
2\Delta\psi-T^{00}\right]  +\frac{m_{g}^{2}}{2}\left[  6\psi^{2}+4\psi
\Delta\pi\right.  \right.  \nonumber\\
&  \left.  \left(  \Delta\pi\pi_{,ik}\pi_{,ik}-\pi_{,ki}\pi_{,ij}\pi
_{,jk}\right)  -2\psi\left(  \pi_{,ik}\pi_{,ik}-2\left(  \Delta\pi\right)
^{2}\right)  \right]  \nonumber\\
&  \left.  +O\left(  \psi^{3},\psi^{2}\phi,\psi^{2}\Delta\pi,\phi\psi\Delta
\pi...\right)  \right\}  \label{20}%
\end{align}
Varying this action with respect to $\phi,$ $\psi$ and $\pi$, and assuming
that $\Delta\pi\ll1$ we obtain the following equations%
\begin{equation}
\Delta\psi=\frac{T^{00}}{2},\text{ \ \ }\Delta\left(  \psi-\phi-m_{g}^{2}%
\pi\right)  -3m_{g}^{2}\psi=0,\label{21}%
\end{equation}%
\begin{equation}
\Delta\psi+\frac{1}{2}\left(  \Delta\pi\pi_{,ik}\right)  _{,ik}+\frac{1}%
{4}\Delta\left(  \pi_{,ik}\pi_{,ik}\right)  -\frac{3}{4}\left(  \pi_{,ij}%
\pi_{,jk}\right)  _{,ik}=0.\label{22}%
\end{equation}
For consistency, we have to include the higher order terms in $\Delta\pi$
because otherwise the first equation in (\ref{21}) would contradict  the
equation (\ref{22}). The reason is that the scalar fields in this case are
always in strong coupling regime. In particular, given $\psi$ which is induced
by the matter source according to Poisson equation and remains unmodified at
all, we obtain from (\ref{22}) the following estimate for induced scalar
fields
\begin{equation}
\partial\partial\pi\sim\Delta\pi\sim\sqrt{\psi}.\label{23}%
\end{equation}
Then considering the spherically symmetric source of mass $M_{0}$ from the
second equation in (\ref{22}) one derives%
\begin{equation}
\psi-\phi\simeq O\left(  1\right)  \psi\left(  \frac{r}{R_{V}}\right)
^{5/2}.\label{24}%
\end{equation}
At distances much smaller than Vainshtein radius $R_{V}=\left(  M_{0}%
/m_{g}^{4}\right)  ^{1/5}$ we have $\psi=\phi$ with high accuracy and thus we
recover General Relativity with corrections which are the same as in the case
of Fierz-Pauli mass term (see \cite{ACM}). However, for $r\gg R_{V}$ the
gravitational potential $\phi$ grows as $r^{3/2},$ while $\psi$ decays exactly
as in Newtonian theory. This is due to the fact that the contribution of the
energy of the field $\pi,$ induced by the external source of the matter,
becomes comparable with the energy of this source at the scales larger than
the Vainshtein radius. To find a solution in this range we have to solve
exactly the complete nonlinear system of equations. However it is obvious that
at distances larger than Vainshtein radius we do not reproduce the results of
massive gravity with the Fierz-Pauli mass term (see \cite{ACM}). For the
realistic graviton mass, Vainshtein radius for the Sun is huge and before we
cross it the contribution of the other mass sources in the universe become
important. Smearing the matter distribution and considering the homogeneous
universe we find that for $m_{g}\simeq H_{0}$, where $H_{0}$ is the present
value of the Hubble constant, the Vainshtein radius is of order of horizon
scale $H_{0}^{-1}.$ Therefore massive gravity with action (\ref{17n}) is in
agreement with experiment. An interesting question that needs investigation is
to determine how General Relativity will be modified on the horizon scale (a
question which could be relevant for the dark energy problem).

\section{How dangerous are ghosts?}

It is clear that the linear ghost around trivial background with $\phi^{A}=0$
is extremely dangerous because it leads to a catastrophic instability of the
vacuum and drastically reduces the lifetimes of the particles. We have shown
in \cite{mukh} how this ghost can be easily avoided. In distinction from it
the nonlinear ghost seems to be unavoidable in all Lorentz invariant versions
of massive gravity. This nonlinear ghost inevitably arises at latest in the
fourth order of perturbation theory on a background which slightly deviates
from the Minkowski space. How dangerous is this ghost? There exist different
opinions on this subject. The main reason why those who think that it is
catastrophic is the integration over the Lorentz boosts in order to insure
Lorentz invariant cutoff. Leaving the question of the need to integrate over
boosts aside we note however that anyway the nonlinear ghost appears only on
the background which deviates from the Minkowski space. In turn this
background selects the preferable coordinate system where we have a Lorentz
violating cutoff on the energy scale below which ghost exists. This cutoff is
the corresponding Vainshtein energy scale, which is extremely low, of order of
$10^{-20}$ $eV$ for the realistic graviton mass. It is clear that the ghost
with such energies is completely harmless from the point of view of agreement
with experiments \cite{CJM}. Therefore we believe that the nonlinear ghost in
any theory of massive gravity is irrelevant. In such case one could wonder if
we can avoid the requirement that the only possible Lorentz invariant graviton
mass term is the Fierz-Pauli one? To answer this question let us consider the
theory with the action%
\begin{equation}
S_{\phi}=\frac{m_{g}^{2}}{8}\int d^{4}x\,\sqrt{-g}\left[  \bar{h}^{2}-\bar
{h}_{B}^{A}\bar{h}_{A}^{B}+\alpha\bar{h}^{2}+O\left(  \bar{h}^{3},...\right)
\right]  .\label{25}%
\end{equation}
It is easy to see that if $\alpha$ is different from zero then already at
quadratic order in the action there appears the term $\alpha\left(  \dot{\chi
}^{0}\right)  ^{2}$ which inevitably leads to a dangerous linear ghost.
Moreover, for $\alpha\sim O\left(  1\right)  $ the Vainshtein scale disappears
in this theory. This can be easily seen if we rewrite equations (31), (39) and
(41) from our previous paper \cite{ACM} taking into account the relevant
contributions from $\alpha\bar{h}^{2}$ term in action (\ref{25})%
\begin{equation}
\Delta\left(  \phi+\psi\right)  +\frac{\alpha}{3\alpha+2}\Delta\left(
\phi-\psi\right)  =T^{00}+m_{g}^{2}\times\left(  ...\right)  ,\label{26}%
\end{equation}%
\begin{equation}
\left(  2\psi-\phi\right)  +\frac{\alpha}{\alpha+1}\left(  \psi+\Delta
\pi\right)  +\partial^{4}\pi^{2}=0,\label{27}%
\end{equation}%
\begin{equation}
\left(  1+2\alpha\right)  \psi+\frac{\left(  3\alpha+2\right)  \left(
\alpha+1\right)  }{2}m_{g}^{2}\pi+\alpha\Delta\pi+\partial^{4}\pi
^{2}=0.\label{28}%
\end{equation}
The nonlinear Vainshtein scale was determined before by the requirement that
in equation (\ref{28}) the linear term in $\pi$ is equal to the last
non-linear term. However, we now have also an extra linear term in this
equation which is always larger than the non-linear term if $\Delta\pi\ll1.$
Hence the non-linear term in this equation is negligible and we always remain
in week coupling regime. By considering the scales for which $k^{2}\gg
m_{g}^{2}$ it follows from (\ref{28}) that
\begin{equation}
\Delta\pi=-\frac{\left(  1+2\alpha\right)  }{\alpha}\psi.\label{30}%
\end{equation}
Substituting this expression in (\ref{27}) we find that up to the leading
order $\psi=\phi$ and hence as follows from (\ref{26}), curiously enough,
General Relativity is restored (at least in the leading approximation) without
having problem with vDVZ discontinuity \cite{dam,zak}. Nevertheless the above
theory is unacceptable because of the linear ghost which exists at all scales
up to the Planckian one.

\section{Conclusions}

We have investigated the problem of the non-linear Boulware-Deser ghost in
massive gravity. In particular, we have used the gravity Higgs mechanism to
study whether the unique theory proposed in \cite{gr,gr1} remains ghost free
away from decoupling regime. Although we have confirmed the result of
\cite{gr,gr1} in decoupling limit, we unfortunately find by explicit
calculations that a nonlinear unremovable ghost reappears in this theory below
Vainshtein energy scale in fourth order of perturbation theory provided away
from the unphysical decoupling limit. At the same time, as was shown in
\cite{mukh,ACM}, the theories considered in \cite{gr,gr1}, can discretely
change the Vainshtein scale within the range $M_{0}^{1/3}m_{g}^{-2/3}%
<R_{V}<M_{0}^{1/5}m_{g}^{-4/5}$. Thus, the claim that massive gravity with
Vainshtein scale $M_{0}^{1/3}m_{g}^{-2/3}$ is ghost free is not confirmed in
the full theory and moreover the nonlinear ghost problem does not seem to be
directly related to the concrete value of the Vainshtein scale.

Higgs massive gravity \cite{mukh,ACM} is equivalent to the formulation in
\cite{gr,gr1} provided one preserves the fake \textquotedblleft Lorentz
invariance\textquotedblright\ in the space of the scalar field configurations.
We have shown that in Higgs gravity, in distinction from \cite{gr,gr1}, the
ghost can be canceled. This, however, can only be done if we abandon the
\textquotedblleft Lorentz invariance\textquotedblright\ in the scalar field
configuration space without violating the fundamental space-time Lorentz
invariance and diffeomorphism invariance. As a result the mass term for the
graviton does not lose its explicit Lorentz invariant form and the massive
graviton inevitably has only two physical degrees of freedom.

To summarize, we have shown that even for the simplest action, which at
leading order reproduces the Fierz-Pauli mass term and ignoring the higher
order terms in $\bar{h}_{B}^{A}$, the Boulware-Deser ghost will arise in third
order of perturbation theory. Moving away from the decoupling limit, while
keeping the contributions of the vector modes in the action, we have
established the existence of the ghost state. We calculated the mass of the
ghost mode $m_{Gh}$ in the short wavelength approximation for perturbations
around some locally Lorentz violating background. Moreover, with strong enough
background fields it is possible to make the negative energy mode as light as
needed within the interval $m_{g}<m_{Gh}<\Lambda_{5}$. However, as was argued
in \cite{ACM}, above the Vainshtein energy scale $\Lambda_{5}$ the scalar
metric perturbations $\psi$ as well as the scalar field perturbations
$\chi^{A}$ are in the strong coupling regime and possess no propagator.
Therefore, the ghost is propagating on the locally nontrivial background only
below the Vainshtein energy scale which for a graviton mass of the order of
the present Hubble scale is extremely low and hence the ghost is harmless.

Further, we have shown that by adding terms of higher order in $\bar{h}%
_{B}^{A}$ to the action with the choice of coefficients corresponding to the
Vainshtein scale $\Lambda=m_{g}^{8/11}$ the nonlinear ghost disappears at the
third order of perturbations. However, away from the decoupling limit the
Boulware-Deser ghost, although harmless, appears at the fourth order of
perturbation theory and cannot be removed by adding higher order terms to the
Lagrangian. This allows us to conclude that the value of the Vainshtein scale
which tells us up to which energy scale a perturbation theory of a given order
is trustable and the presence of the nonlinear ghost in the theory are two
separate issues which do not have to be correlated.

We have argued that because of diffeomorphism invariance of the variables
$\bar{h}_{B}^{A}$ appropriate counterterms which violate the "fake Lorentz
invariance"can be added to the Lagrangian so that the action takes the form
(\ref{17n}). This cancels the undesired negative energy mode. The propagators
for the scalar and vector modes of the massive graviton vanish as a result of
which the action (\ref{17n}) will describe a massive graviton with two degrees
of freedom.

\begin{acknowledgement}
We are grateful to G. Dvali, G. Gabadadze, L. Berezhiani and A. Gruzinov for
discussions. \textit{The work of AHC is supported in part by the National
Science Foundation grant 0854779. L.A and V.M. are supported by TRR 33
\textquotedblleft The Dark Universe\textquotedblright\ and the Cluster of
Excellence EXC 153 \textquotedblleft Origin and Structure of the
Universe\textquotedblright.}
\end{acknowledgement}


\end{document}